\documentclass[9pt,twocolumn,twoside]{article} 
\usepackage{graphicx}
\usepackage{textcomp}
\newcommand{\JournalTitle}[1]{\textbf{#1}}

\title{Generation and characterization of few-pulse attosecond pulse trains at
  100\,kHz repetition rate}
\date{}

\begin{document}
\twocolumn[
\maketitle
\vspace{-8ex}
Mikhail Osolodkov$^1$, Federico J. Furch$^1$, Felix Schell$^1$, Peter
\v{S}u\v{s}njar$^1$, Fabio Cavalcante$^2$, Carmen S. Menoni$^2$, Claus
P. Schulz$^1$, Tobias Witting$^{1,*}$, Marc J. J. Vrakking$^1$
\\
\textit{$^1$Max Born Institute for Nonlinear Optics and Short Pulse
  Spectroscopy, Max-Born-Strasse 2a, 12489 Berlin, Germany}
\\
\textit{$^2$Department of Electrical and Computer Engineering, Colorado
  State University, Fort Collins, CO 80523, USA}
\\
\textit{$^*$Corresponding author: tobias.witting@mbi-berlin.de}
\\

\textit{ Many experiments in attosecond science will benefit from
  attosecond pulses at high repetition rates with sufficient photon
  flux for pump-probe experiments. We use 7\,fs, 800\,nm pulses from a
  non-collinear optical parametric chirped pulse amplification
  (NOPCPA) laser system to generate few-pulse attosecond pulse trains
  (APTs) in the extreme ultraviolet (XUV) at a repetition rate of
  100\,kHz. The pulse trains have been fully characterized by
  recording FROG-CRAB (Frequency-Resolved Optical Gating for Complete
  Reconstruction of Attosecond Bursts) traces with a velocity map
  imaging spectrometer.}
\vspace*{4ex}
]

Advances in the field of attosecond science are strongly related to
advances of ultrashort laser sources. High energy near infrared (NIR)
pulses with durations on the femtosecond timescale allow the
generation of attosecond pulses in the extreme ultraviolet (XUV)
through the process of high-order harmonic generation
(HHG)~\cite{Krausz}. Combined with phase-locked short NIR pulses in a
pump-probe configuration, attosecond XUV pulses can be used for
attosecond timescale spectroscopy
experiments~\cite{lepine_attosecond_2014}. The process under
investigation can be studied by measuring the absorption of the
attosecond pulse or by detecting momentum distributions of
photoelectrons or ions resulting from the interaction of the sample
with a sequence of laser pulses. A more sophisticated approach than
the latter is the coincidence detection of all charged particles
following the interaction utilizing reaction microscopes or COLTRIMS
(COLd Target Recoil Ion Momentum Spectroscopy)
detectors~\cite{Ullrich}. Detection in coincidence allows measurement
of the correlated three-dimensional momentum distributions of all
charged particles resulting from the interaction. That information
provides access to the photoelectron momentum distributions in the
recoil frame in the case of a dissociating molecule. Previously
applied to strong field ionization
experiments~\cite{Feuerstein,Eremina}, the coincidence detection
technique has recently also been implemented for attosecond
spectroscopy measurements. The combination of attosecond pump-probe
spectroscopy with coincidence detection has made the spatio-temporal
reconstruction of the photoionization process in a CO
molecule~\cite{vos_orientation-dependent_2018} possible, and has been
used to characterize the influence of electron-nuclear coupled
dynamics on the phase of the entangled electron-nuclear wave packet
during the dissociative ionization of H$_2$
molecules~\cite{cattaneo_attosecond_2018}.

In order to perform coincidence detection measurements, a single
ionization event restriction should be fulfilled, i.e. at most one
ionized atom or molecule per laser shot. On the other hand, the number
of detected events should be high enough to achieve good
statistics. Together these two conditions result in the demand for
high repetition rate laser systems with a sufficiently high pulse
energy to generate XUV pulses. In recent years significant development
of high repetition rate ($\gg$ 10\,kHz) high energy laser sources
enabled XUV generation at high repetition rates~\cite{Hadrich}. A
number of groups have reported the generation of high-order harmonics
at high repetition rate \cite{Harth,Gonzalez,Lorek,Krebs}.  So far,
with the exception
of~\cite{hammerland_rabbitt_2019,hammerland_effect_2019}, who reported
RABBITT experiments at 100\,kHz using a fiber laser post-compressed to
40\,fs, no pump-probe experiments or pulse characterization have been
reported. In particular, there have not been any reports yet
demonstrating the complete characterization of attosecond pulses
generated using a high repetition rate, OPCPA driver, which is
important given concerns that have existed in the past about
spatio-temporal couplings in such
systems~\cite{giree_numerical_2017,Witting_spatiotemporalOPCPA,Harth}.
In this Letter we report on the generation of few-pulse XUV attosecond
pulse trains (APTs) in the 15 to 40\,eV energy range with a flux of
$>10^6$ photons per shot at a repetition rate of 100\,kHz. We fully
characterize the APTs using FROG-CRAB (Frequency-Resolved Optical
Gating for Complete Reconstruction of Attosecond
Bursts)~\cite{Mairesse}. The APTs are accompanied by phase-locked
$<10$\,fs NIR pulses for pump-probe experiments.

\begin{figure*}[htbp]
\centering
\fbox{\includegraphics[width=\linewidth]{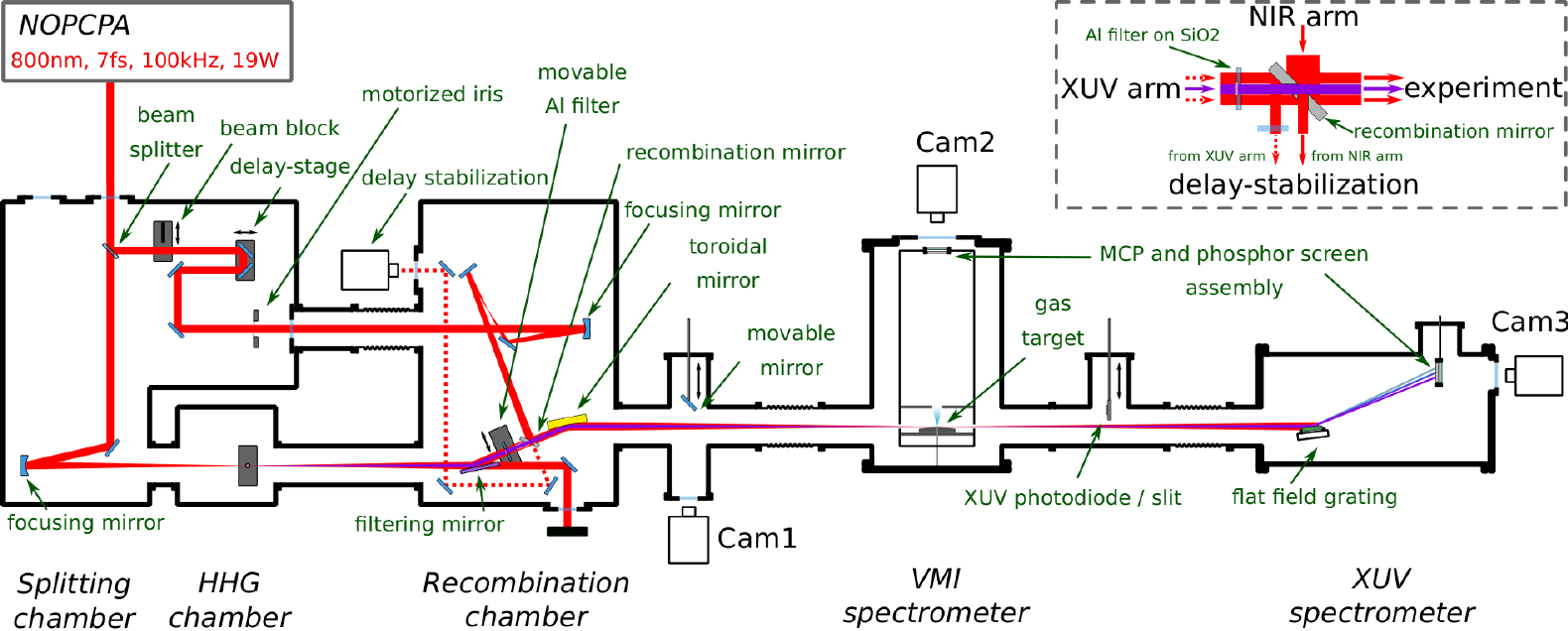}}
\caption{Schematic top view of the experimental setup. The NIR and XUV
  beams are shown in red and violet respectively. \textbf{Inset:}
  schematic view of the recombination mirror. For a detailed
  description please refer to the text.}
\label{fig:beamline}
\end{figure*}

A sketch of the experimental setup is shown in
Fig.~\ref{fig:beamline}. A noncollinear optical parametric chirped
pulse amplification (NOPCPA) system is utilized as the drive laser for
the experiment. It delivers up to 190\,\textmu{}J, 7\,fs pulses
centered at 800\,nm with a repetition rate of 100\,kHz
\cite{Furch,Witting_spatiotemporalOPCPA}. The attosecond pump-probe
setup consists of three separate interconnected vacuum chambers. In
the \textit{Splitting chamber} the NIR s-polarized pulses from the
NOPCPA system are split into two beams with an 80/20 ratio using a
beam splitter with a flat response over the spectral range 500-1100
nm. The majority (80\,$\%$) of the incoming power is transmitted
through the beam splitter and then used for the generation of
high-order harmonics. In what follows we are going to refer to this
arm of the interferometer as the \emph{XUV arm} and to the other one
as the \emph{NIR arm}. In the \textit{Splitting chamber} the high
power NIR beam is focused by a spherical mirror ($f = 50$\,cm) into a
2\,mm long gas cell positioned inside the \textit{HHG chamber}. The
NIR pulse compression is optimized for efficient high-order harmonic
generation. The NIR peak intensity in the focus is on the order of
200\,TW/cm$^2$. After high-order harmonic generation the high power
NIR beam co-propagating with the XUV beam is filtered out in two steps
in the \textit{Recombination chamber}. First, the two-color beam is
reflected by a specially designed dichroic mirror.
The mirror has a Ta$_2$O$_5$/SiO$_2$ broadband antireflection coating
for s- and p- polarization of the NIR beam topped by a 200\,nm thick
layer of SiO$_2$. At a grazing incidence angle of 75 degrees the
coating reflects the XUV beam with an average reflectivity of 45$\%$
in the 20-60\,eV spectral range whereas only 20$\%$ of the NIR light
is reflected, independent of the polarization.
The rest of the NIR beam is transmitted by the filtering mirror and
then sent to an external beam dump. In the second filtering step the
remaining 20$\%$ of the NIR beam is blocked by a 200\,nm thick
aluminium filter mounted on a 1\,mm thick fused silica plate.
In the NIR arm of the interferometer, a translation stage with
nanometer precision (SmarAct SLC-1750-S-HV) is used to control the
delay between the pulses propagating in the two arms of the
interferometer. A motorized iris (SmarAct SID-5714) installed in the
NIR arm is used to control the NIR intensity delivered to the
experiment by changing the size of the clear aperture.
The beams from the XUV and NIR arms are recombined utilizing a
recombination mirror shown schematically in the inset of
Fig.~\ref{fig:beamline}. The recombination mirror has two 5\,mm
diameter drilled holes. They are oriented at 90\,degrees with respect
to each other and at 45\,degrees to the mirror surface. The size of
the clear aperture of the Al filter is the same as the size of the
holes in the recombination mirror. The XUV beam transmitted through
the Al filter propagates through one of the holes in the mirror. Part
of the NIR beam from the NIR arm is reflected on the recombination
mirror around the hole, resulting in an annular beam that is then
recombined with the XUV beam and co-propagates towards the
experiment. In the NIR arm we generate a focus in the equivalent plane
to the HHG target. A gold-coated toroidal mirror forms a $1:1$ image
(object distance 1250\,mm) of the two planes in the interaction region
of the detection system, resulting in spatially overlapped foci of the
XUV and NIR beams.

\begin{figure}[htbp]
\centering
\fbox{\includegraphics[width=\linewidth]{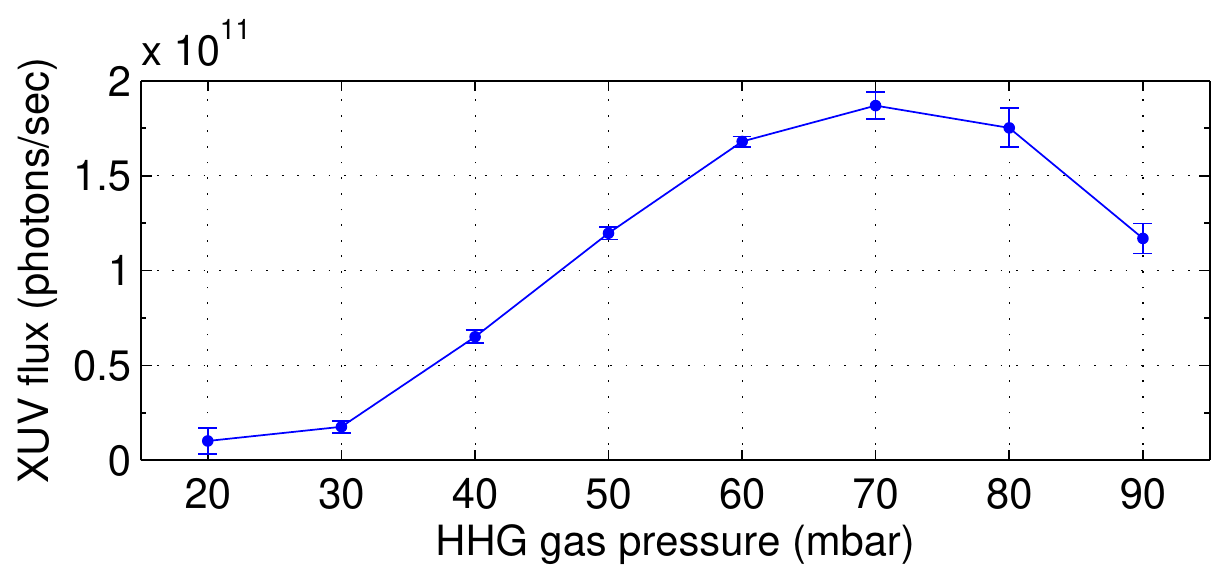}}
\caption{XUV photon flux on target as a function of the Kr pressure in
  the HHG target gas cell.}
\label{fig:flux}
\end{figure}

\begin{figure*}[htbp]
\centering
\fbox{\includegraphics[width=0.8\linewidth]{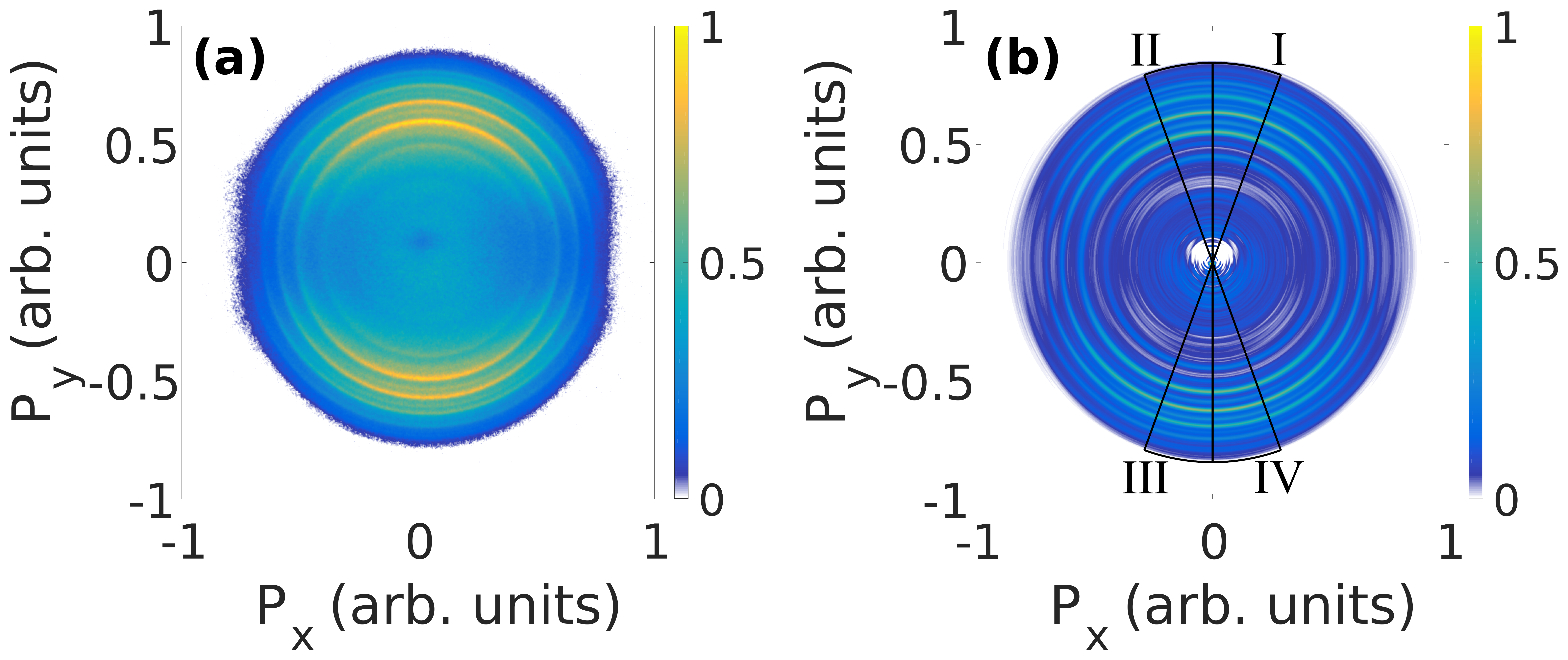}}
\caption{Two-color XUV+NIR VMI images with XUV and NIR pulses at
  temporal overlap. Raw data normalized to the maximum value (a), and
  and slice through the 3-dimensional retrieved momentum distribution
  (b). The NIR field-induced sidebands are visible along the laser
  polarization axis (parallel to $p_y$). The four sections in (b)
  indicate the angular ranges used to create the four FROG-CRAB
  matrices employed for the four independent pulse reconstructions.}
\label{fig:vmi}
\end{figure*}

To enable attosecond pump-probe experiments the XUV- and NIR laser
pulses from the pump-probe interferometer need to have a stable
time-delay. We use a combination of passive and active stabilization
schemes. The entire interferometer is built inside the vacuum system
onto optical breadboards connected to an optical table and isolated
from the chambers by bellows. The chambers are vibration-isolated from
the optical table. For active stabilization we use the second output
port of the recombination mirror. The beam layout is sketched in the
inset of Fig.~\ref{fig:beamline}.  We use the residual NIR from the
XUV arm in combination with the transmitted NIR beam through the
recombination mirror. The central part of the beam from the NIR arm of
the interferometer propagates through the second hole in the
recombination mirror. Since the Al filter in the XUV arm is mounted on
a fused silica plate and the size of the NIR beam incident on the
Al-filter is larger than its diameter, the outer part of the beam is
transmitted through the glass mount, while the central part is blocked
by the filter. This annular beam is reflected by the back-side of the
recombination mirror and then used together with the transmitted part
of the NIR arm beam for delay-stabilization. We intentionally
unbalance the glass amount in the two NIR beams (see inset
Fig.~\ref{fig:beamline}) to introduce a time-delay of
$\approx{}500$\,fs. The resulting spectral interference fringes carry
a $\cos{(\Delta\phi+\omega\tau)}$ modulation and are recorded by a
home-made spectrometer which is equipped with a fast USB\,3 camera
(Basler acA2040-90umNIR). The phase-shift is evaluated by a PC with a
fast fourier transform (FFT)-based algorithm and a feedback signal is
created controlling the delay-stage. Using a small region of interest
on the camera chip we reach acquisition loop speeds of up to
1\,kHz. However, we restrict the write speed to the piezo stage to
$<250$\,Hz to avoid resonances. This way pump-probe scans over 100s of
fs can be performed with attosecond phase-lock. The delay stability is
around 50\,as rms over hours.

After the recombination chamber overlapping XUV and NIR pulses are
focused onto a gas target in a velocity map imaging spectrometer
(VMI). The sample atoms are ionized by the XUV beam and the resulting
three dimensional photoelectron momentum distributions are projected
onto a two-dimensional detector (micro channel plate (MCP) and
phosphor screen assembly) by a set of two electrodes forming an
electrostatic lens. A 50\,$\mu m$ nozzle creating the gas target is
incorporated into the repeller electrode in order to increase the gas
density in the interaction region~\cite{Ghafur}. The signal on the MCP
and phosphor screen assembly is recorded using a CMOS (complementary
metal–oxide–semiconductor) camera (Basler acA2040-90umNIR). An XUV
spectrometer is positioned behind the VMI. It allows detection of the
spatially resolved XUV spectrum during the experiment. The XUV
spectrometer consists of a flat field grating (Hitachi) mounted on a
motorized stage and an MCP and phosphor screen assembly detector. The
signal on the detector is imaged by CCD (charge-coupled device) camera
(BASLER scA1400-17gm) installed outside of the vacuum chamber.

In the present experiment high-order harmonics were generated in
krypton gas. The XUV photon flux available for experiments was
measured using an XUV photodiode (Opto diode AXUV100G) installed on a
movable mount between the VMI and XUV spectrometer. The XUV photon
flux as a function of the Kr pressure in the gas cell is shown in
Fig.~\ref{fig:flux}. The pump-probe measurements were performed with
70\,mbar in the gas cell, which corresponds to an XUV photon flux of
more than $1.8\times{}10^{11}$ photons per second, or
$1.8\times{}10^{6}$ photons per shot in the interaction region of the
experiment.

\begin{figure}[htbp]
\centering
\fbox{\includegraphics[width=\linewidth]{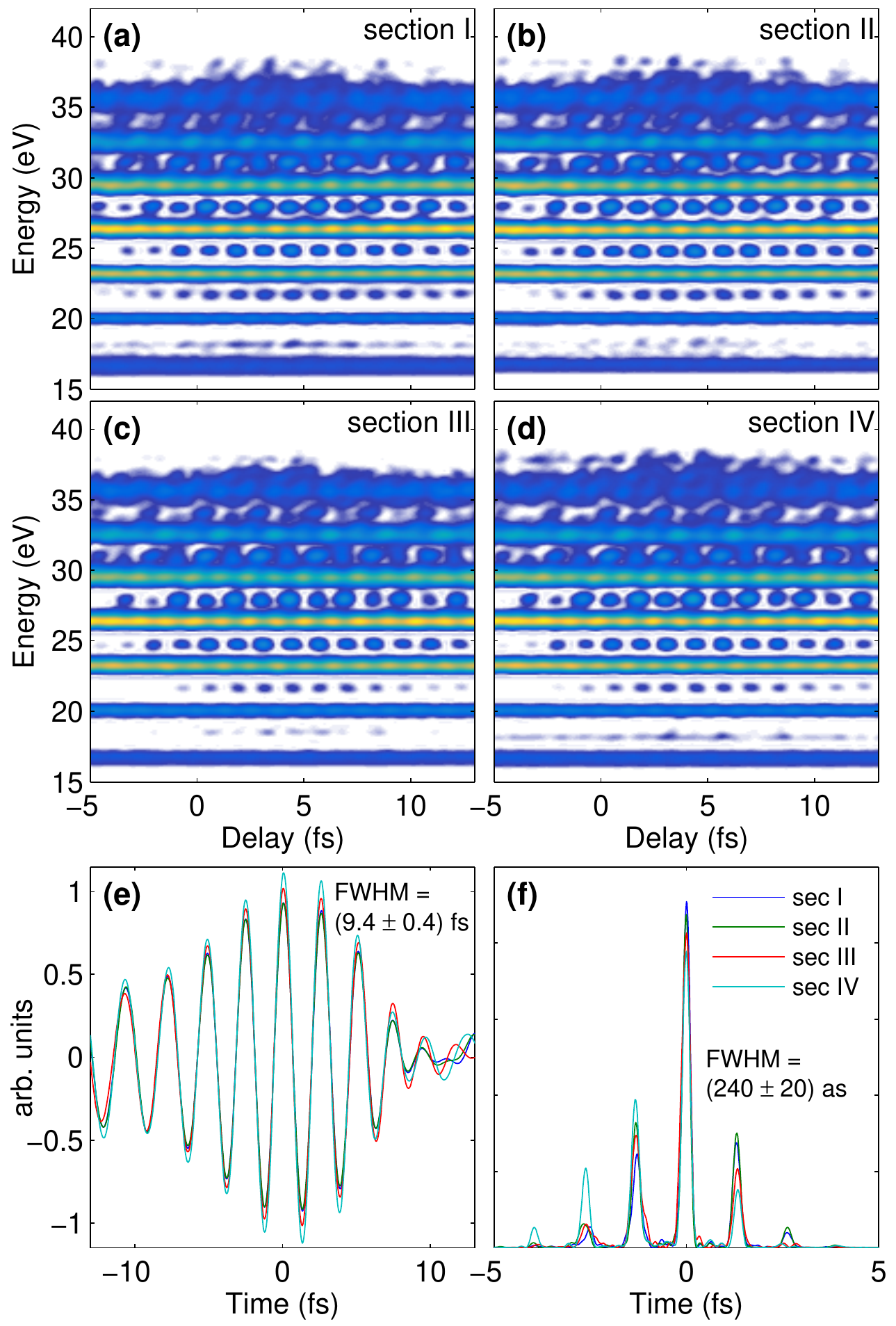}}
\caption{FROG-CRAB traces created by angular integration of areas I,
  II, III, and IV as indicated in Fig.~\ref{fig:vmi}(b) (a) to
  (d). Reconstructed NIR pulse electric fields (e) and reconstructed
  XUV attosecond pulse trains for each section (f).}
\label{fig:crab}
\end{figure}

In the present pulse characterization experiment argon was used as a
gas target. The Ar atoms were ionized by the XUV single photon
absorption in the presence of a NIR field of intensity
3\,TW/cm$^{2}$. Since the three-dimensional photoelectron momentum
distributions have cylindrical symmetry along the laser polarization
axis (along $p_y$ in Fig.~\ref{fig:vmi}) they can be unambiguously
retrieved from two-dimensional projections on a plane parallel to the
axis of symmetry using an Abel inversion. Measured VMI images were
inverted, using an inversion method based on fitting the measured
projections of angular distributions in polar coordinates with
Legendre polynomials~\cite{Garcia}. VMI images were acquired for a
pump-probe delay range of up to 40\,fs with 100\,as steps.
The photoelectron momentum distributions were integrated over four
different angle ranges of 20\,degrees with respect to the laser
polarization axis. The angular ranges of integration are shown as
areas I, II, III and IV, in the retrieved three-dimensional momentum
distribution in Fig.~\ref{fig:vmi}(b). Kinetic energy distributions
were subsequently obtained by interpolating the integrated momentum
distributions over an equally spaced kinetic energy axis.

In Figure \ref{fig:crab}(a) to (d) the resulting photoelectron kinetic
energy distributions are plotted as a function of the delay between
the NIR and XUV pulses. The energy axis is shown as photon energy,
i.e. the ionization potential of argon (15.6\,eV) is added to the
energies of the detected photoelectrons.
From the resulting photoelectron kinetic energy distributions as a
function of the delay both the XUV and NIR pulses can be
reconstructed. As our laser pulses of 7\,fs generate APTs with only a
few pulses, we are in a regime between long-pulse RABBITT where one
can assume a periodic APT and streaking with isolated attosecond
pulses. This manifests itself in the recorded FROG-CRAB traces
(Fig.~\ref{fig:crab}) as non-sidebandlike modulation especially around
the higher energy harmonics. The structure and strength of the
modulations are dependent on the NIR delay and therefore intensity, as
opposed to regular same-strength modulations seen for a 10s~of~fs
modulating pulse. Near the highest harmonics also the `harmonics' are
modulated in a streaking-like manner.

Therefore, a RABBITT-type analysis of fitting the sideband
oscillations is not appropriate in our case. In order to retrieve the
XUV and NIR pulses, we employ a time-domain ptychography
algorithm~\cite{spangenberg_ptychographic_2015,Lucchini,Witting_timedomainptychography}. This
iterative algorithm enables the reconstruction of the complex electric
fields of both, the unknown XUV, and the dressing NIR pulses. It was
demonstrated to be more robust and converging faster than other well
known reconstruction algorithms, such as the PCGPA (principal
component generalized projections algorithm) or LSGPA (least squares
generalized projections algorithm)~\cite{Lucchini}. We run the
time-domain ptychography algorithm in its extended ptychographic
iterative engine (ePIE)
implementation~\cite{Lucchini,Witting_timedomainptychography}
independently for the four traces from areas I, II, III, and IV. The
time-delay axis was kept as acquired in the experiment with
$\delta\tau = 0.1$\,fs and the photoelectron spectra were re-sampled
onto a $2^{12}$ pixel grid ($\delta{}t = 20$\,as,
$\delta\hbar\omega = 50$\,meV). The ePIE algorithm retrieves the
amplitudes and phases of the electron wavepacket from which we
retrieve the complex electric field of the APT by subtracting the
dipole transition phases~\cite{mauritsson2005}. The reconstructed NIR
electric fields are shown in Fig.~\ref{fig:crab}(e). The NIR pulses
had an intensity FWHM duration of $9.4\pm0.4$\,fs. The pulse duration
in the NIR arm is slightly longer than in the XUV arm, as the two arms
are not perfectly dispersion-balanced. The 1\,mm window between
splitting and recombination chambers (see. Fig.~\ref{fig:beamline})
does not exactly compensate the dispersion of the 1\,mm beamsplitter
mounted at 45\,deg. The reconstructed APTs (shown in
Fig.~\ref{fig:crab}(f)) consist of six individual attosecond pulses
with one dominating main pulse. The duration of the central attosecond
pulse is $240\pm20$\,as.

In conclusion, we have demonstrated the generation of attosecond pulse
trains at 100\,kHz repetition rate. We fully characterized the APTs
using the FROG-CRAB scheme with a time-domain ptychography retrieval
algorithm. Henceforth, the attosecond beamline reported here is going
to be used for two-color XUV-NIR pump-probe spectroscopy measurements
with coincidence detection in molecules.

\bigskip
\textbf{Funding.} European Union Horizon 2020 program Laserlab Europe
(638585); European Union Horizon 2020 Marie Curie ITN project ASPIRE
(674960). CSM acknowledges support of grant DoD ONR  N00014-17-1-2536.

The authors thank R. Peslin, A. Loudovici, and Ch. Reiter for
technical support.

\end{document}